# Design and simulation of GaSe hybrid photonic waveguides on a β-Ga$_2$O$_3$ platform

Md. Nakibul Islam[1], Md. Islahur Rahman Ebon[1], Sergey A. Degtyarev[2,3] and Jaker Hossain[1*]

[1]*Photonics & Advanced Materials Laboratory, Department of Electrical and Electronic Engineering, University of Rajshahi, Rajshahi 6205, Bangladesh.*

[2]*Samara National Research University, Moskovskoye Shosse 34, Samara, 443086, Russia*

[3]*Image Processing Systems Institute, FRC "Kurchatov Institute", Molodogvardeiskaya Str. 151, Samara 443001, Russia.*

**Abstract**

This study explores the potential of a novel material platform combining Gallium Selenide (GaSe) with a β-Ga$_2$O$_3$ substrate and Al$_2$O$_3$ cladding for advanced photonic waveguides. The numerical investigation of the optical performance of GaSe-based waveguides focuses on key parameters such as power confinement factor (PCF), propagation loss, and effective mode index for both of the modes of Transverse Electric (TE) as well as Transverse Magnetic (TM). The results demonstrate that the GaSe/Al$_2$O$_3$/β-Ga$_2$O$_3$ waveguide exhibits excellent optical confinement, with PCF values exceeding 96% for both the TE and TM modes at a wavelength of 1.55 μm. The waveguide also shows low propagation losses, particularly for the TM mode, making it suitable for long-wavelength applications. Furthermore, the study highlights the impact of core dimensions and cladding height on waveguide performance, providing insights into optimizing the design for minimal loss and maximal efficiency that demonstrates the potential of III-VI compound-based waveguides. This treatise also focuses on the bending losses by varying the wavelength and bending radius for both of the TE and TM mode of the GaSe-based waveguide. The waveguide shows the reflecting loss of -25.08 dB and transmission loss of -5.97 dB for the bending radius of 2 μm and wavelength of 1.55 μm. This work underscores the potential of GaSe-based waveguides on a β-Ga$_2$O$_3$ platform for next-generation photonic integrated circuits.

**Keywords:** GaSe, β-Ga$_2$O$_3$, Al$_2$O$_3$, rectangular waveguide, COMSOL Multiphysics.

## 1. Introduction

Photonic integrated circuits (PICs) have gained significant interest over traditional discrete photonic components due to their superior performance in high-speed signal processing and data transmission [1]. In the recent time, silicon is used as a core material for the PICs owing to its well-established complementary metal oxide semiconductor (CMOS)-compatible device manufacturing processes that offer building of large-scale PICs with a bit rise in cost [2-3]. Despite the incredible success, the silicon PICs have some drawbacks because of the indirect bandgap (BG) of silicon that impede by fabricating a really compacted silicon-based photo-electrical platform [2]. In addition, the non-direct BG of silicon consequences in the break off wavelength at 1.1 μm, and the multi-photon absorption (TPA) phenomenon results in a higher loss at the telecommunication wavelengths [4]. Moreover, absence of the Pockels effect in silicon hinders in achieving expected speed in light modulation and efficiency [5]. The advancement of PICs is also relying on choice of the material which is situated in the compatible platforms that are susceptible in intern photons in the nanometer-scale. Therefore, searching new compounds with ultra-low loss and high electro-optic (EO) coefficients is immensely required in advancing PICs. As a number of compound stage, for example III-V semiconducting compounds [6-7], lithium niobate (LiNbO$_3$, LN) [8-9], silicon nitride (Si$_3$N$_4$) [10], indium phosphide (InP) [11] and diamond [12-13] have been widely studied for PICs. Herein, we proposed an III-VI β-Ga$_2$O$_3$ compound platform for an III-VI GaSe compound for photonic waveguides.

β-Ga$_2$O$_3$, like gallium nitride (GaN) and silicon carbide (SiC) semiconductors, is being considered as a viable alternative material for various applications owing to its large bandgap and relatively inert nature [14]. Ga$_2$O$_3$ has the bandgap of ~4.9 eV and extensively used in photocatalysts, ultraviolet (UV) photodetectors, solar cells, gas sensors, luminescent devices [15-17], metal-oxide-semiconductor field effect transistors (MOSFETs) [14, 18], and Schottky barrier diodes [19]. Recently, it has been used as a low loss waveguide material in the UV–NIR spectra [1]. Ga$_2$O$_3$ thin films have been synthesized by various methods, such as vacuum thermal evaporation, spray pyrolysis, sol-gel, pulsed laser deposition (PLD), molecular beam epitaxy (MBE), radio frequency magnetron sputtering, and CVD [15].

In addition, recently, group III-VI van der Waals (vdW) compounds, have gained much attraction due to their extraordinary optical and photonics properties [20]. Especially, GaSe exhibits thickness dependent optoelectronic properties and its bandgap may be largely tuned which is flexible for the used in photoelectric and ultrafast photonics devices [21]. GaSe is reported to have four phases, namely, ε, β, γ, and δ, where ε-phase shows excellent optoelectronic and photonic properties [20]. GaSe crystal and thin films have been deposited by, so far, chemical vapor transfer (CVT), thermal evaporation, Chemical Bath Deposition (CBD), etc. [20, 22-23].

Aluminum oxide ($Al_2O_3$) is an enthralling cladding candidate for photonic waveguides in the group of III-VI. $Al_2O_3$ is a widely recognized material used in various optical components due to its superb transparency across the UV to NIR spectrum and its relatively high refractive index compared to traditional $SiO_2$ cladding [24-25]. In recent years, $Al_2O_3$ thin films have been widely utilized across various applications, including optical waveguides, amplifiers, optical lenses and windows, refractory and antireflection coatings and lasers [26]. To achieve high-quality $Al_2O_3$ thin films for these diverse applications, various deposition processes have been introduced, inclusive of chemical vapor deposition, thermal evaporation, electron beam evaporation, spray pyrolysis, anodization, the sol-gel process, aerosol-jet deposition, atomic layer deposition and magnetron sputtering. Surrounded by these methods, chemical vapor deposition is widely applied for producing thin films with high-quality crystal structure and with excellent stoichiometry as well as constancy. Moreover, its requirement for a substrate with higher temperature (approximately 500 °C) is often undesirable for many applications [27]. Moreover, it shows great potential for integrated optical applications, like in the telecommunication window, $Al_2O_3$ is used to operate in a low-loss waveguide (0.21 dB/cm) [28].

Waveguides serve as fundamental building blocks in photonic integrated circuits (PICs), where the confinement factor and waveguide loss are pivotal in determining the overall efficiency and functionality of the system.

This research investigates the potential of a GaSe-based straight waveguide structure, incorporating an $Al_2O_3$ cladding layer and a β-$Ga_2O_3$ substrate, through detailed numerical simulations. In this study, it is seen that, the III-VI group is a promising and propitious group for the design of the rectangular or hybrid waveguides. To optimize performance, the GaSe/$Al_2O_3$/β-$Ga_2O_3$ waveguide was systematically analyzed by varying critical parameters, including operating wavelength, core width, and core height. The findings highlight the waveguide's promising characteristics for advanced photonic applications.

## 2. Structure of the device and computation

The design and simulations of the Gallium Selenide (GaSe) waveguide has been performed using COMSOL Multiphysics 6.2, a finite-element-based software widely utilized for electromagnetic wave analysis [29]. The Electromagnetic Waves, Frequency Domain (ewfd) module has employed to numerically solve Maxwell's equations using the finite-element method and compute key optical parameters, including Intensity Mode Profile, Power Confinement Factor (PCF), Loss, and Refractive Mode Index for TE and TM modes. The Effective Mode Index ($n_{eff}$) in a photonic waveguide measures the phase velocity of a guiding mode compared to the light speed in vacuum. It shows the influence of the waveguide's structure and materials on light propagation. Mathematically, it's expressed in the equation 1 [30].

$$n_{eff} = \frac{\beta}{k_0} \quad \ldots\ldots\ldots\ldots\ldots\ldots\ldots\ldots\ldots(1)$$

Where propagation constant of the mode is denoted by the $\beta$ and the free-space wavenumber is $k_0 = \frac{2\pi}{\lambda_0}$, with $\lambda_0$ being the wavelength in vacuum. The COMSOL Multiphysics solves different equations to solve or perform the waveguide simulation. The equation 2 is the frequency-domain form of the wave equation derived from Maxwell's equations is known as Vector Helmholtz Equation or simply the Electromagnetic Wave Equation. It describes the propagation of the electric field $E$ in a medium with relative permittivity $\epsilon_r$. The term $k_0$ represents the wave number in free-space [31].

$$\nabla \times (\nabla \times \boldsymbol{E}) - k_0^2 \epsilon_r E = 0 \ldots\ldots\ldots\ldots (2)$$

Also, the Complex Propagation Constant equation is displayed in equation 3 [32].

$$\alpha = j\beta + \delta_z = -\lambda \ldots\ldots\ldots\ldots\ldots\ldots.. (3)$$

Where, the propagation constant $\alpha$ is a complex quantity that describes how a wave propagates and attenuates in a medium, $j\beta$ is the imaginary part of the propagation constant, where $\beta$ is the phase constant, $\delta_z$ is the real part of the propagation constant, representing attenuation (or loss) in the $z$-direction and $\lambda$ is the wavelength of the wave. The Modal Field Representation or Waveguide Mode Representation equation is given in equation 4 and 5 [33].

$$E(x, y, z) = E(x, y)e^{-\alpha z} \ldots\ldots\ldots\ldots\ldots.. (4)$$

$$H(x, y, z) = H(x, y)e^{-\alpha z} \ldots\ldots\ldots\ldots\ldots.. (5)$$

This equation represents the electric field in a waveguide or similar structure, where the field can be separated into a transverse component $E(x, y)$ for TE mode and $H(x, y)$ for the TM mode and a propagation term $e^{-\alpha z}$. The term $e^{-\alpha z}$ describes how the field decays and changes phase as it propagates in the $z$-direction.

Also, the Scattering Boundary Condition or Radiation Boundary Condition equation is expressed as equation 6 [29].

$$\boldsymbol{n} \times (\nabla \times \boldsymbol{E}) - jkn \times (\boldsymbol{E} \times \boldsymbol{n}) = 0 \ldots\ldots\ldots\ldots (6)$$

It is derived from the Sommerfeld radiation condition, which ensures that waves radiate outward to infinity. Where, $n$ is the unit normal vector to the boundary and $k$ is the wave number in the medium. Perfect Electric Conductor (PEC) Boundary Condition is shown in equation 7 [34].

$$\boldsymbol{n} \times \boldsymbol{E} = 0 \ldots\ldots\ldots\ldots\ldots (7)$$

This condition enforces that the tangential component of the electric field $\boldsymbol{E}$ is zero at the boundary of a perfect conductor and $\mathbf{n}$ is the unit normal vector to the boundary. It is derived from the fact that the electric field inside a perfect conductor is zero, and the tangential component must vanish at the surface. These equations are fundamental in the study of electromagnetics and are widely used in computational electromagnetics software like COMSOL Multiphysics for simulating wave propagation, scattering, and boundary effects.

The waveguide structure has been modeled with a GaSe core, $\beta$-$Ga_2O_3$ substrate, and $Al_2O_3$ cladding, ensuring an accurate representation of material properties and optical confinement behavior. The Table 1 shows the various parameters of the rectangular waveguide.

**Table 1. Parameters of basic structure of the waveguide**

| Parameters | Value |
|---|---|
| Optimal wavelength | 1.55 μm |
| Width of the waveguide | 4.0 μm |
| Height of the waveguide | 3.0 μm |
| Width of the core | 1.2 μm |
| Height of the core | 0.8 μm |
| Width of the substrate | 4.0 μm |
| Height of the substrate | 1.0 μm |
| Bending radius | 2.0 μm |
| Refractive index of GaSe | 2.7576 [35] |
| Refractive index of $\beta$-Ga$_2$O$_3$ | 1.9095 [36] |
| Refractive index of Al$_2$O$_3$ | 1.7462 [37] |
| Laser input power | 1 W |
| CTE of GaSe | Linear $10.8 \times 10^{-6}$ °C$^{-1}$ [38] |
| CTE of $\beta$-Ga$_2$O$_3$ | $4.2 \times 10^{-6}$ °C$^{-1}$ [39] |
| CTE of Al$_2$O$_3$ | $8.2 \times 10^{-6}$ °C$^{-1}$ [40] |

## 2.1. Structure of the Waveguide

Figure 1(a) depicts a planner waveguide structure composed of a $\beta$-Ga$_2$O$_3$ substrate, a GaSe core, an Al$_2$O$_3$ cladding, and these are situated in a silicon (Si) wafer, each playing a critical role in its optical performance. The $\beta$-Ga$_2$O$_3$ substrate is providing mechanical support and influencing the overall waveguide properties. The GaSe core, which is a fundamental and most prominent layer. With its high refractive index, it ensures optimal confinement of the optical field, supporting efficient light propagation. The Al$_2$O$_3$ cladding surrounding the core with a lower refractive index ensures the conditions for total internal reflection thereby decreasing optical propagation losses. The Si layer at the bottom adds structural strength and robustness.

Figures 1(b) and 1(c) depict the mode intensity profiles for the TE and TM modes, which show the optical power confinement within the waveguide at a wavelength of 1.55 μm. Figure 1(b) portrays the TE mode, where the front view displays the horizontal intensity profile of the waveguide, indicating strong optical confinement with minimum field penetration into the cladding. The gradual intensity decay at the edges shows efficient light propagation with minimal optical power loss within the waveguide. The top view confirms stable mode propagation along the waveguide, showing a consistent intensity profile with minimal disruptions. Achieving a Power Confinement Factor (PCF) of 96.40% for TE mode, while the Effective Refractive Index is 2.5681, confirming efficient confinement and stable optical transmission. Figure 1(c) depicts the intensity mode profile of the TM mode of the waveguide at a wavelength of 1.55 μm. The PCF for the TM mode is 96.15%, which is slightly lower than the TE mode's PCF, while the effective refractive index for this particular mode is 2.5881, slightly higher than that of the TE mode at a particular wavelength of 1.55 μm.

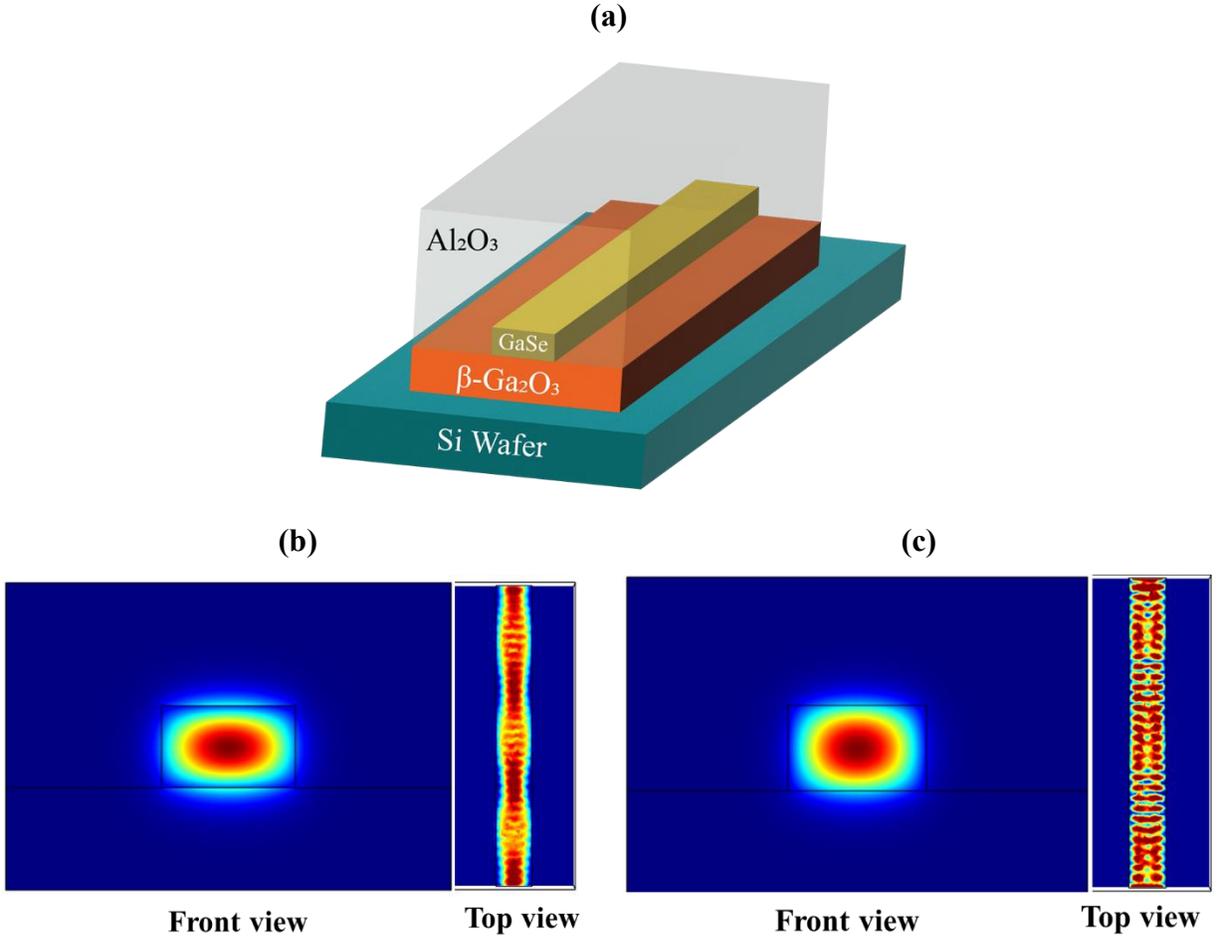

**Figure 1.** Schematic diagram of the (a) rectangular waveguide, and mode intensity profiles: (b) front and top views of the waveguide in TE mode and (c) front and top views of the waveguide in TM mode.

## 3. Results and Discussion

### 3.1. PCF, Loss, and Effective Mode Index Analysis with Varying Wavelengths

Figure 2(a) illustrates the PCF for both of the TE and TM modes as a operation of wavelength. Both modes exhibit a similar trend, PCF initially remains constant at 100% in both TE and TM modes and then decreases with increasing wavelength. However, the mode TM consistently shows superior PCF liken to the mode of TE across the entire wavelength. The PCF remains 90% for both TE and TM modes at a wavelength of 2 μm. As the wavelength increases, the PCF rapidly decreases to 75% for the TM mode and 70% for the TE mode at a wavelength of 3.15 μm. The effective refractive index exhibits a reduction as the wavelength increases, a behavior that plays a significant role in inducing variations within the PCF. This wavelength-dependent decrease in the effective refractive index is a key factor influencing the dispersion properties and modal characteristics of the PCF, thereby affecting its optical performance and guiding mechanisms [41]. At longer wavelengths, the optical field tends to extend further beyond the core region, resulting in reduced interaction between the propagating wave and the waveguide's core. This phenomenon can be explained through the principles of modal dispersion. As the wavelength increases, a greater portion of the optical mode extends into the cladding material, reducing the intensity of the field confined within the core. This behavior is fundamental to understanding the wavelength-dependent properties of PCFs and their modal characteristics [42-43].

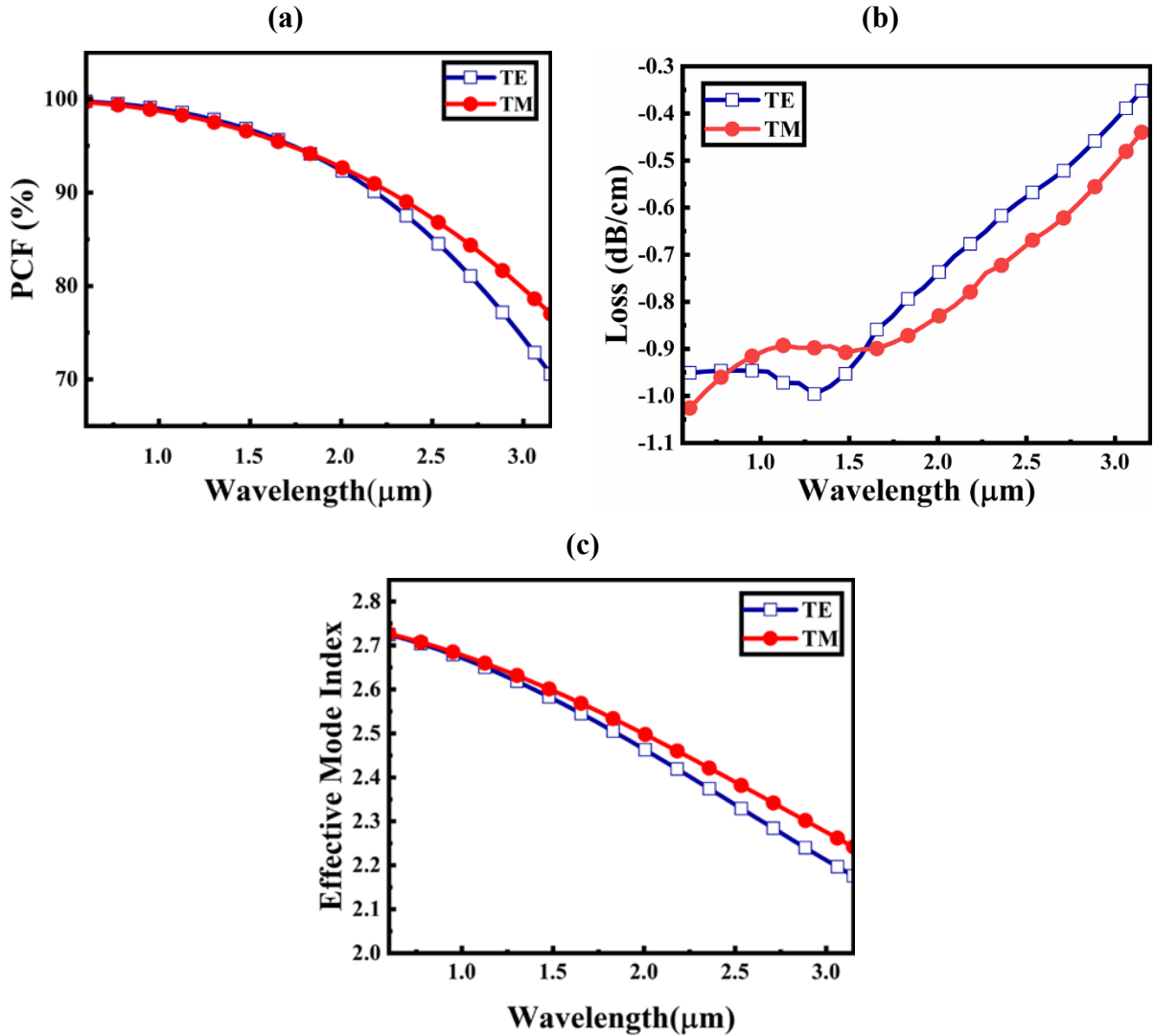

**Figure 2.** The alteration of (a) PCF, (b) Loss, and (c) Effective Mode Index with wavelength.

Figure 2(b) shows the losses for TE and TM modes of the waveguide, measured in dB/cm, across a wavelength range of 0.1 μm to 3.25 μm. At shorter wavelengths, both modes exhibit low and relatively similar losses, approximately around -1.0 dB/cm. As wavelengths increase beyond 1.5 μm, the loss for the TE mode begins to rise more rapidly than for the TM mode. This difference becomes more significant at longer wavelengths, with the TE mode reaching a loss near -0.2 dB/cm at 3.0 μm, while the TM mode remains around -0.4 dB/cm. This trend indicates that the TM mode achieves better propagation efficiency with lower attenuation, highlighting its suitability for longer-wavelength applications. As the wavelength increases, the PCF decreases and the loss increases. Because the loss is the phenomena where the power is not confined properly. As the wavelength escalates, a greater portion of the optical mode extends into the cladding material, diminishing the degree of confinement within the core [42-43]. This similar trend is located in some previous reports [44].

Figure 2(c) depicts the change of the effective mode index (EMI) with wavelength for modes of TE as well as TM across wavelengths from 0.1 μm to 3.15 μm. The degree of light confinement within the waveguide, which is indicated by the effective mode index, decreases as the wavelength increases for both TE and TM modes. This phenomenon occurs due to the inverse relationship between the effective refractive index and the wavelength of light. [45]. This indicates the waveguide becomes less effective at confining light at longer wavelengths. TM mode consistently exhibits a slightly higher EMI than the TE mode across the entire wavelength range. The mode index which is effective of both modes are close at shorter wavelengths. As the wavelength approaches 3.15 μm, the Effective Mode Index of both TE and TM modes decreases, and the gap between TE and TM widens slightly. This inconsistency may occur because the TM situation is significantly responsive to inaccuracies in vertical layer properties—such as thickness, angle, and material anisotropy—than to those in horizontal layers [46-47].

## 3.2. PCF, Loss, and Effective Mode Index Analysis with Varying Height of the Core

Figure 3(a) demonstrates the variation of PCF with respect to the height of the core material for the modes of TE and TM. The height of the core varies from 0.2 μm to 1.15 μm, and the PCF behavior is analyzed for both TE and TM modes. As the height of the core increases, the PCF for both TE and TM modes also increases, indicating improved confinement of optical power within the GaSe core. Initially, when the height of the GaSe core is around 0.2 μm, the PCF is significantly lower for both TE and TM modes, particularly for TE mode. TM mode consistently exhibits upper PCF contrast to the TE mode across all the heights of the core variation. The TM mode exhibits a consistently higher PCF compared to the TE mode across all core heights. This indicates that the TM mode offers better power confinement in the waveguide than the TE mode. The height of the core of the primary waveguide was initially set to 0.8 μm, where the PCF for both TE and TM modes is observed to reach 96.5%. As the height of the GaSe core approaches 1 μm, the increase in PCF becomes slower, and values for both TE and TM modes near saturation. When the core height is at its maximum of 1.15 μm, the power confinement factor for both of the modes reaches 98%. A waveguide with a larger cross-sectional dimension enhances the light internment within the high-index core region, and as the dimensions become significantly large, the confinement factor approaches to 100% [48].

Figure 3(b) depicts the variation in loss (dB/cm) with the height of the core for TE and TM modes. The height of the core varies from 0.2 μm to 1.15 μm, and the loss of the GaSe waveguide is observed for both TE and TM modes. As the height of the core increases, the loss for both TE and TM modes generally decreases, suggesting better optical power confinement and reduced propagation loss at a larger height of the core. However, the behavior is not uniform across different heights of the core. For the TE mode, the loss initially fluctuates at smaller core heights, reaching a peak around 0.4 μm before experiencing a steady decline beyond 0.5 μm. As the core height increases, the TE mode loss stabilizes around -0.9 dB/cm near 1.0 μm, indicating better efficiency of the waveguide at larger core heights. In contrast, the TM mode exhibits a constant loss of approximately -5 dB/cm at lower core heights, peaking around 0.5 μm, before experiencing a rapid decrease in loss until the core height reaches 1 μm. Beyond 1 μm, the loss stabilizes around -8.5 dB/cm, indicating a significant improvement in waveguide efficiency as the core height increases. However, the waveguide is initially designed with a core height of 0.8 μm, where both TE and TM mode losses are observed to be approximately -0.9 dB/cm. Thus, increasing the core height enhances waveguide performance, with the TM mode showing more stability and lower losses than the TE mode. These results highlight the importance of optimizing core height to achieve minimal propagation losses and maximize optical confinement efficiency. The loss is the phenomena where the power is not confined properly. As the core height escalates, a greater portion of the optical mode extends into the core material, uprising the degree of confinement within the core [42-43]. That is why, the loss is seen to be reduced with the variation of the height of the core material [48].

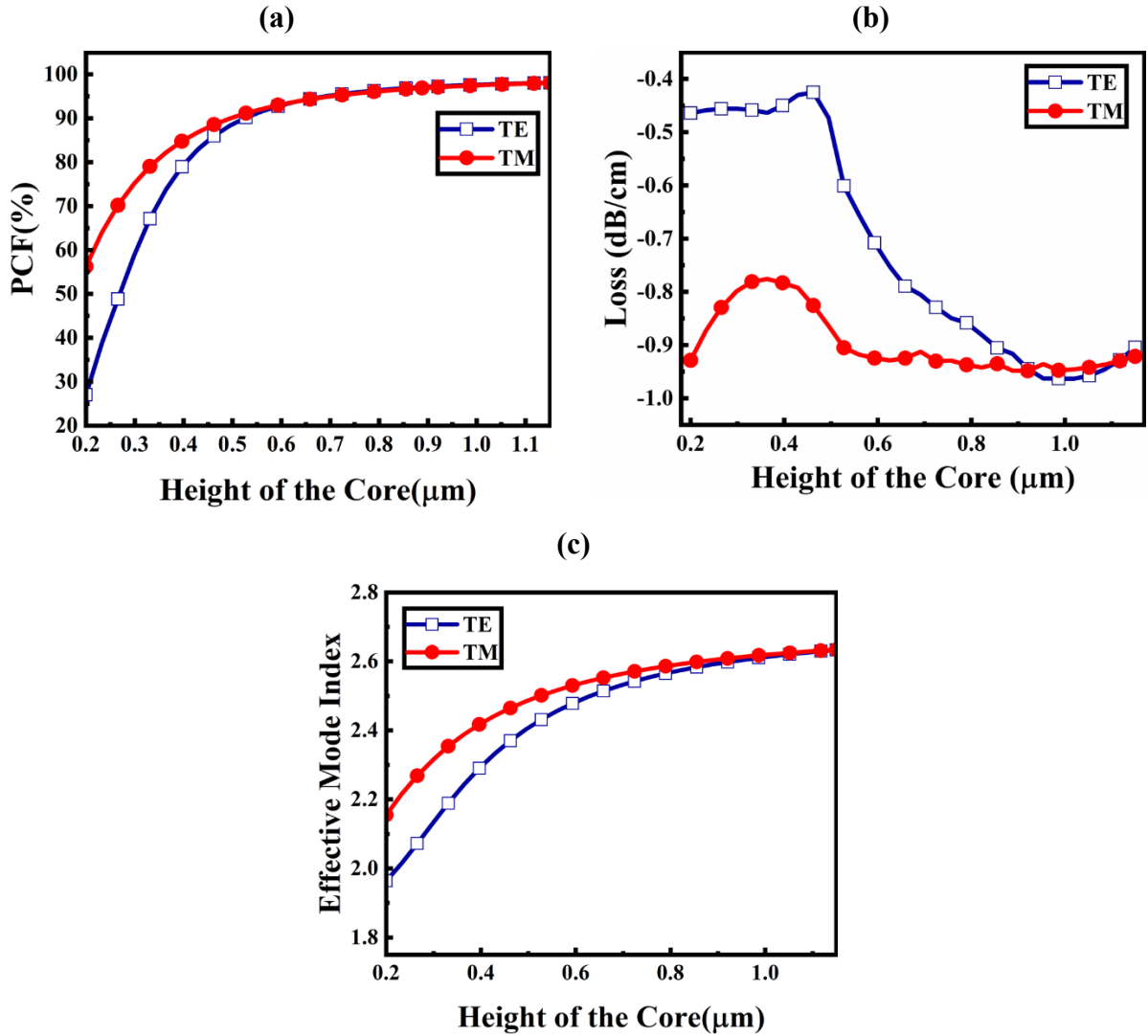

**Figure 3.** The variation of (a) PCF, (b) Loss, and (c) Effective Mode Index with Height of the Core.

Figure 3(c) illustrates the variation of the effective mode index for the height of the core in both TE and TM modes. The height of the core varies from 0.2 μm to 1.15 μm, and the EMI is analyzed for TE and TM modes. As the height of the core increases, the effective mode index for both TE and TM modes increases, indicating better power confinement of the optical field within the waveguide. At a lower height of the core, around 0.2 μm, the effective mode index for both TE and TM modes is relatively low, with the TM mode exhibiting a slightly higher value compared to the TE mode. This indicates that the TM mode has better confinement characteristics even at smaller core heights. As the height of the core approaches 1.15 μm, the effective mode index for both modes gradually approach to the saturation. The TM mode consistently exhibits a superior effective mode index compared to the TE mode throughout the entire range of core heights. This implies that the TM is more effective at confining the optical field within the waveguide compared to the TE mode. As the dimensions of the GaSe slabs are enlarged, a greater portion of light becomes confined within the higher-index core slabs, leading to an increase in the effective indices [48]. The mode index for TE and TM situation increases, with a slight widening of the gap between the two. This discrepancy may arise because TM situation is considerably more responsive to inaccuracies in vertical layer properties—such as thickness, angle, and material anisotropy—compared to those in horizontal layers [46-47].

### 3.3. PCF, Loss, and Effective Mode Index Analysis with Varying Width of the Core

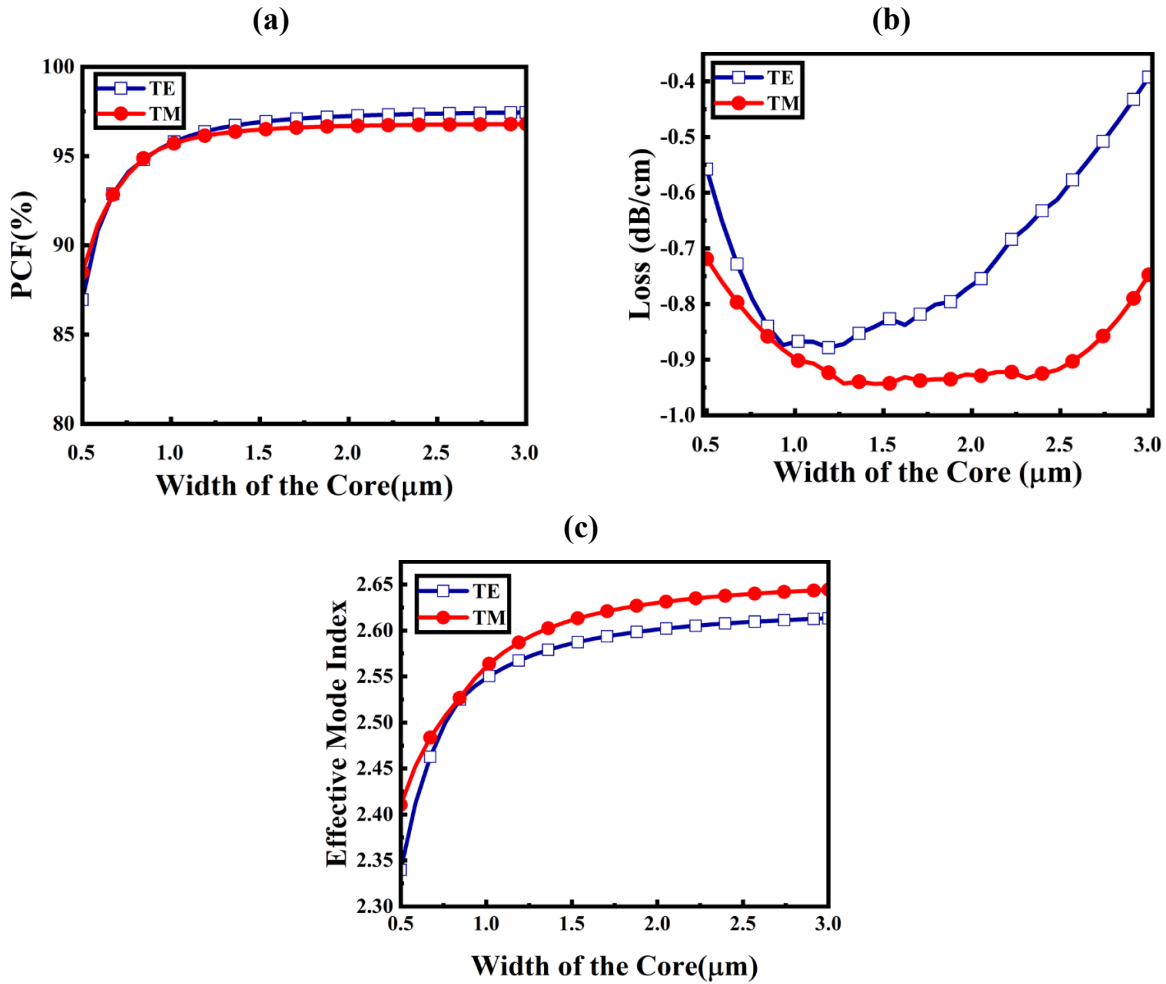

**Figure 4.** The change of (a) PCF, (b) Loss, and (c) Effective Mode Index with width of the core

Figure 4(a) illustrates PCF as a dependance of the width of the core for both TE and TM modes. The width of the core is varied from 0.5 µm to 3.0 µm, and the PCF is observed for both TE and TM modes. Initially, with a core width of 0.5 µm, the PCF is observed to be around 85% for both situations. As the core width increases, a significant improvement in PCF is observed, with both TE and TM modes reaching above 95% around the 1.0 µm of core width. Beyond 1.0 µm, the PCF remains nearly constant, indicating a saturation point in the optical confinement performance. Notably, the TM mode consistently displays a higher PCF relative to the TE mode across various core widths. By the time the core width reaches 3.0 µm, the PCF for TE and TM modes reaches a stable level of approximately 97%, showing minimal dependency on further increase of the width of the core. Setting the width of the GaSe core of the waveguide initially to 1.2 µm, we observed that the PCF for both TE and TM modes reached 96%. This finding highlights that optimal confinement of optical power within the waveguide is achieved at around 1.0 µm. Beyond this width, the improvement becomes negligible. A waveguide with an increased cross-sectional dimension improves the confinement of light within the high-index core region, and as the dimensions grow substantially, the confinement factor approaches 100% [48]. As the core width increases, a greater portion of the optical mode extends into the GaSe core material, enhances the intensity of the field confined within the core [42-43, 48].

Figure 4(b) illustrates the variation in propagation loss (dB/cm) with the width of the core for both TE and TM modes, where the core width is varied from 0.5 µm to 3.0 µm. At the narrower width of the core, particularly at 0.5 µm, the TE mode experiences a higher propagation loss of about -0.5 dB/cm, whereas the TM mode experiences a lower loss of approximately -0.7 dB/cm. Increasing the core width reduces propagation losses for both TE and TM modes, with the TE mode reaching its minimum loss around 1.0 µm, while the TM mode achieves its lowest loss near 1.5 µm**.** The TE mode shows a higher loss compared to the TM mode over the entire range. Initially, the width of the core of the primary waveguide is set to 1.2 µm, showing around -0.95 dB/cm loss for TM and -8.5 dB/cm loss for TE mode. As the core width approaches 3.0 µm, the TE mode loss rises significantly, reaching approximately -0.4 dB/cm, while the TM mode loss remains more stable in comparison, reaching around -0.75 dB/cm. As the core width increases, a larger fraction of the optical mode propagates within the core material, enhancing the degree of light confinement in the core [42-43]. Consequently, the loss decreases as the core width material is varied [48]. But this zigzag behavior is also located in some previous reports [44, 49].

Figure 4(c) illustrates the variation in the effective mode index for the width of the core for both TE and TM modes. The core width varies from 0.5 μm to 3.0 μm and the EMI for both modes are analyzed. The Effective Mode Index increases as the core width increases for both TE and TM modes, indicating better optical confinement and reduced propagation loss. At a core width of 0.5 μm, the Effective Mode Index for both TE and TM modes are initially lower, around 2.34 and 2.41, respectively, indicating weaker confinement at a smaller core width. At a width of approximately 1.0 μm, a sharp rise in the effective mode index is noted, especially for the TM mode, which surpasses the TE mode consistently. After 1.0 μm, the Effective Mode Index growth rate decreases for both TE and TM mode, approaching saturation as the core width reaches 3.0 μm. When the core width is 3.0 μm, the Effective Mode Index reaches approximately 2.64 for TM part and 2.61 for the TE part. TM situation exhibits a sublime EMI than the TE mode throughout all core widths, suggesting better confinement and stability for TM mode. The saturation trend at wider widths suggests that further increases in the core width result in minimal gains for the EMI. TE as well as TM mode indices exhibit an upward trend, accompanied by a slight expansion in the gap between them. This divergence may be attributed to the TM mode's heightened sensitivity to variations in vertical layer properties—such as thickness, angular deviations, and material anisotropy—relative to those in the horizontal layers [46-47]. Because of the dimensions of the GaSe slabs increase, a larger fraction of light is confined within the higher-index core slabs, resulting in a rise in the effective mode indices [48-49].

### 3.4. Impact of Cladding Height Variation on Vertical Cut-Line Profiles

Figure 5 illustrates the effect of cladding height variation on the vertical cut-line. The electric field changes with the y-coordinate, and the field is analyzed for both TE and TM modes. Figure 5(a) shows the visual representation of the 2D cut line of the waveguide.

Figure 5(b) and Figure 5(c) show the changes in the electric field to the y-coordinate for TE and TM modes respectively. The height of the waveguide was varied from 1.9 μm to 3.0 μm to investigate its effect on the electric field distribution along the y-coordinate. The electric field consistently exhibited a peak intensity at approximately 1.45 μm, indicating strong confinement within the core region across all cladding heights. The GaSe core is positioned between 1 μm and 1.8 µm along the y-coordinate. The profiles displayed a rapid decline in intensity beyond the core boundaries, signifying minimal field penetration into the cladding. The electric field intensity for TM part exhibits supreme values contrast to the TE section. Specifically, the peak intensity for TM mode reaches approximately 48 V/m, while for TE mode, it is about 35 V/m, both occurring around 1.45 µm along the y-coordinate. The maximum electric field intensity was slightly decreased by the increasing cladding height, but the overall distribution remained mostly unchanged. This uniformity highlights that the core structure primarily dictates the electric field confinement, while the cladding provides sufficient isolation without significantly affecting the light propagation.

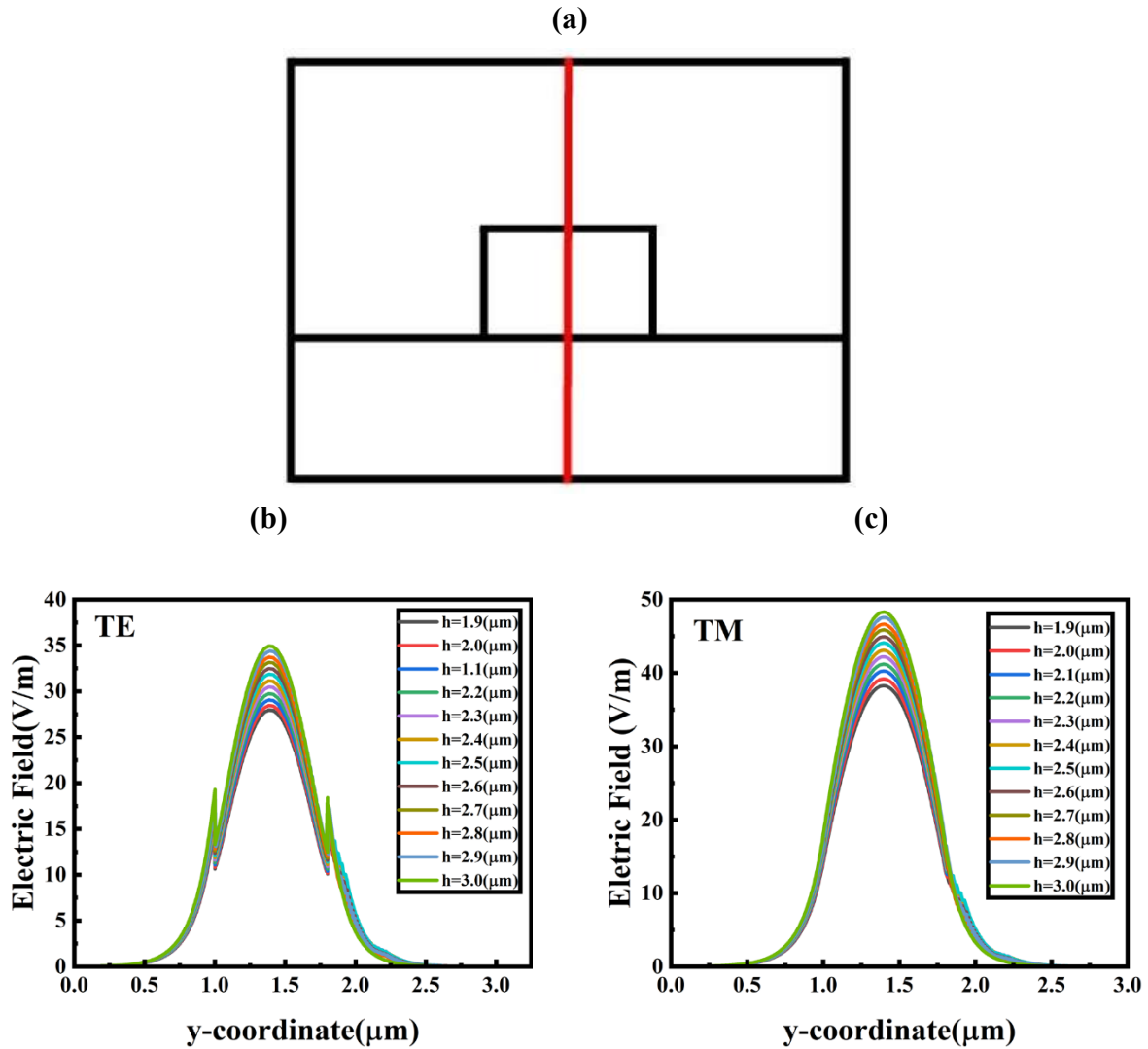

**Figure 5.** The Variation of the Electric Field with the y-coordinate (a) 2D Cut Line, (b) TE Mode, and (c) TM Mode.

These findings demonstrate that variations in cladding height within the studied range have a minimal effect on the confinement and stability of the two modes. Meantime, when the cladding layer is lower, a significant electric field distribution widen has been observed to the $Al_2O_3$ region, while the effect is lower evident within core, because of the r cladding refractive index which is comparatively lower. Also, two additional peaks have appeared on either side of the main peak, resulting from the abrupt refractive index change at the core-cladding interface. These unwanted peaks are absent in the TM because the distinct polarization orientations of the TE as well as TM modes. Moreover, when the direction of cross-sectional rotation has been occurred by the 90°, the peaks are manifested in TM mode rather than TE part [50].

## 3.5. Impact of Cladding Width Variation on Horizontal Cut-Line Profiles

Figure 6 illustrates the electric field intensity distribution along the x-coordinate of the waveguide for various widths, ranging from 1.3 μm to 5.0 μm, depicting both TE and TM modes. Figure 6(a) represents the cut line profile along the x-coordinate of the waveguide, which is essential for understanding the distribution of electric field intensity across different waveguide widths.

Figure 6(b) and Figure 6(c) highlight the electric field intensity distribution along the x-axis for the TE and TM modes, respectively. The electric field intensity peaks are observed to shift as the waveguide width increases for both TE and TM modes. The electric field intensity for Transverse Magnetic mode exhibits bigger values on the comparison of the TE portion. Specifically, the peak intensity for TM mode reaches approximately 54 V/m, while for TE mode, it is about 39 V/m, both occurring around 2.5 μm along the x-coordinate. The electric field intensity is low in the cladding, peaks in the GaSe core, and decreases again upon re-entering the cladding for both TE and TM modes.TM mode exhibits more complex behavior, with oscillatory features and multiple peaks becoming evident as the width increases. The TE mode, in contrast, maintains a more uniform and symmetric profile, even for wider waveguides. Additionally, two additional peaks have appeared on either side of the main peak, resulting from the sudden change of refractive index at the cladding and core junction. These unwanted peaks are absent in the TM because the distinct polarization orientations of the TE as well as TM modes. Moreover, when the direction of cross-sectional rotation has been occurred by the 90°, the peaks are manifested in TM mode rather than TE part [50]. However, this improvement is relatively modest compared to the enhancement achievable with high-index-contrast slot waveguide structures [51].

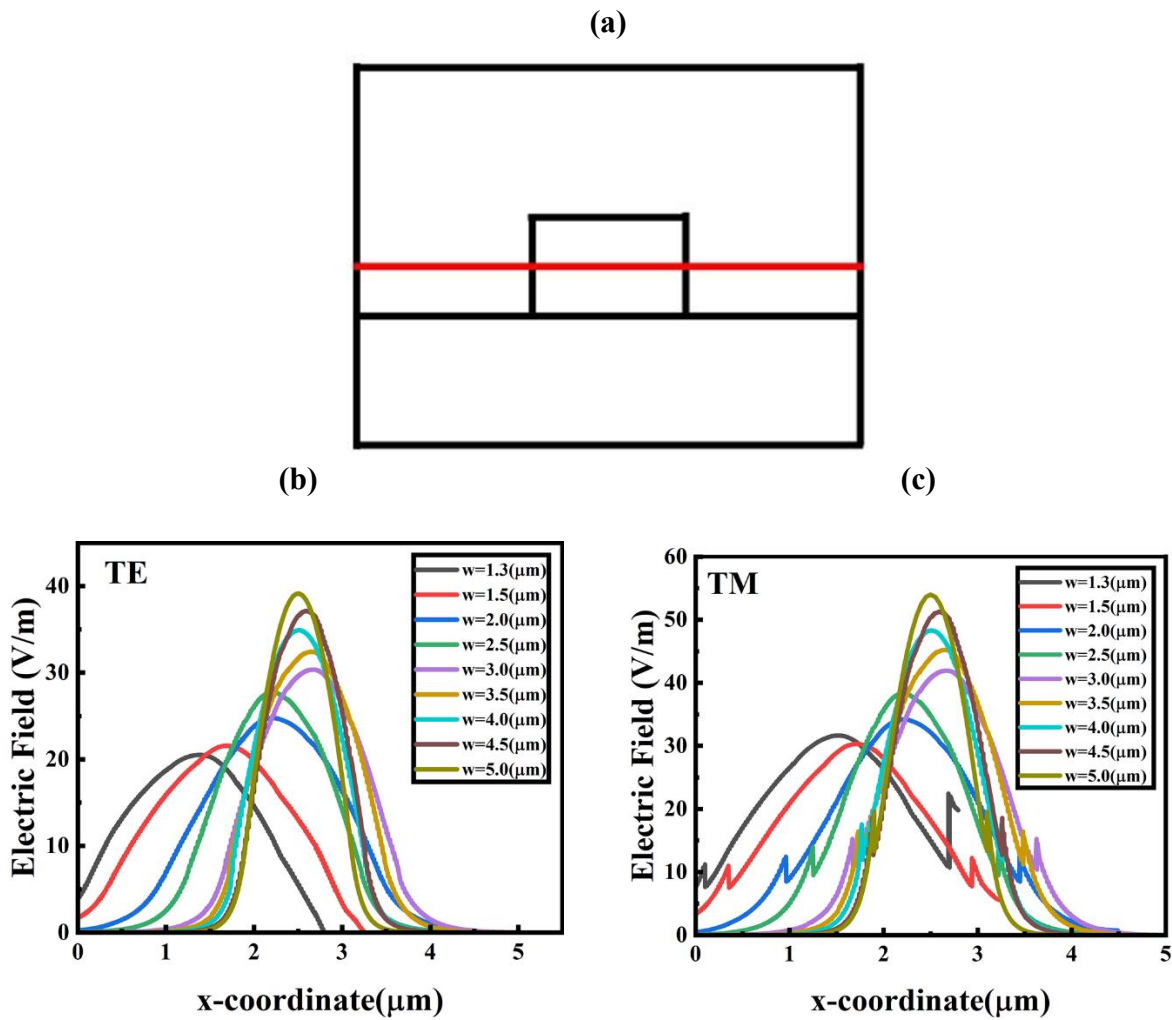

**Figure 6.** The Variation of Electric Field with a length of x-coordinate (a) 2D Cut Line, (b) TE Mode, and (c) TM Mode.

### 3.6 Bending Loss Analysis at Fixed Radius across Wavelengths

The performance of waveguides in optical communication and integrated photonics is heavily influenced by bending loss. When a waveguide undergoes bending, radiation loss occurs due to the mismatch in mode confinement, leading to energy dissipation [52]. This loss is considerably affected by the bending radius, and operational wavelength [53]. This study analyzes the bending loss behavior of a waveguide with a fixed bending radius of 2 μm, where the wavelength is varied from 1.0 μm to 3.0 μm. The focus is on both TE and TM modes, assessing their transmission (S21) and reflection (S11) losses. Notably, 1.55 μm is identified as an important operational wavelength due to its widespread use in optical communication systems [54]. Figure 7(a) shows the schematic view of bending waveguide, indicating the input port, output port and bending radius. Also, Figure 7(b) displays the top view of the bending waveguide after propagation in TE mode.

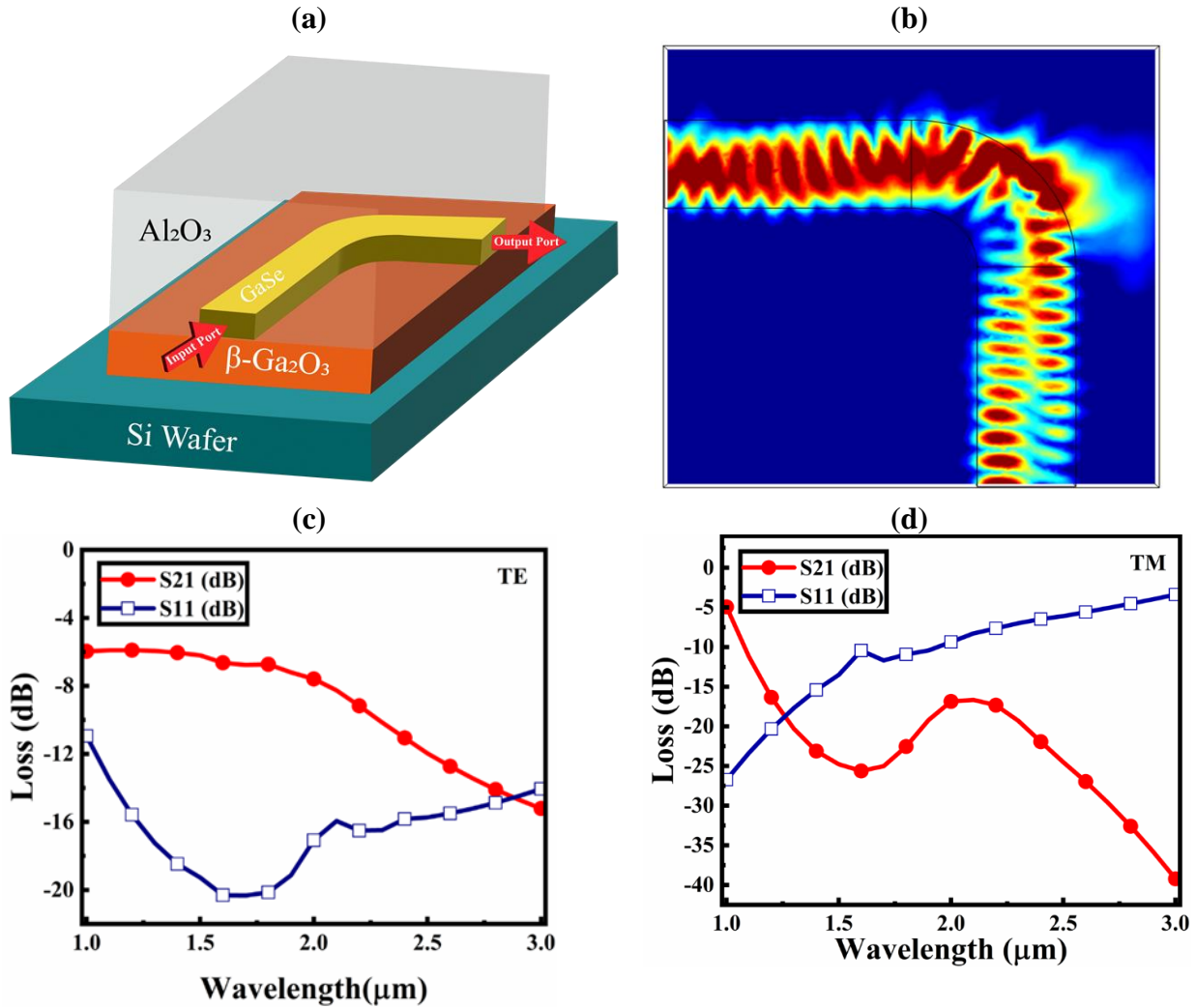

**Figure 7.** (a) Schematic view of bending waveguide, (b) Mode profiling after propagation of bending waveguide, Transmission (S21) and Reflection (S11) losses in the H-bend Waveguide (c) TE mode, and (d) TM mode.

Figure 7(c) presents the transmission (S21) and reflection (S11) losses in an H-bend waveguide for the TE mode. At shorter wavelengths, the transmission loss (S21) remains relatively stable, starting at approximately -8 dB at 1.0 μm. As The wavelength increases beyond 1.5 μm leads to a gradual rise in loss, reaching nearly -16 dB at 3.0 µm. This increase indicates that longer wavelengths experience higher transmission loss in the H-bend waveguide. In contrast, the reflection loss (S11) follows a different trend. It starts at approximately -12 dB at 1.0 µm and decreases, reaching a minimum of nearly -18 dB around 1.5–1.6 µm. Beyond this point, S11 begins to increase again, indicating stronger reflections at higher wavelengths. The observed dip in reflection loss at 1.55 µm suggests that this wavelength exhibits optimal performance, where transmission is relatively high and reflections are minimized. The predominant source of loss arises from the mode mismatch occurring at the interface between the straight and bending waveguides, where the optical modes fail to align perfectly, leading to significant energy dissipation and reduced transmission efficiency [55]. The losses for the bending part escalate significantly at higher wavelengths, also their dependency on the wavelength frequently exhibits pronounced oscillations due to the interference of the reflection of light at the outer coating surface as well as cladding boundary [56].

Figure 7(d) displays the transmission (S21) and reflection (S11) losses observed in an H-bend waveguide operating in TM mode. The transmission loss (S21) undergoes a sharp decline from approximately -5 dB at 1.0 µm to nearly -28 dB at 1.5 µm. A slight recovery is observed around 1.7–2.0 µm, where S21 momentarily increases, but it resumes its downward trend, reaching approximately -40 dB at 3.0 µm. This severe transmission loss at longer wavelengths suggests that the TM mode suffers from higher bending loss, likely due to weaker mode confinement. In contrast, the initial reflection loss (S11) for the TM mode is around -25 dB at 1 µm, and reflection loss increases as the wavelength increases. Unlike the TE mode, where an optimum wavelength exists, the TM mode does not show a specific spectral region of minimal loss. Instead, the results indicate that TM polarization is significantly more vulnerable to bending, making it less favorable for applications requiring low-loss transmission across a wide wavelength range. The losses for the bending portion escalate significantly at higher wavelengths, also their dependency on the wavelength frequently exhibits pronounced oscillations due to the interference of the reflection of light at the outer coating surface as well as cladding boundary [56]. Furthermore, the losses for TE polarization are consistently lower compared to those for TM polarization, as evidenced in both Figure 8 (c) and (d). This discrepancy can be ascribed to the inherent variations in the refractive index between the tendancy of in-plane and out-of-plane, which influence the propagation characteristics and confinement of the respective polarization modes, thereby resulting in differing loss magnitudes [50].

### 3.7 Impact of Radius Variation on Bending Loss at Different Wavelengths

Figure 8 presents the bending loss (dB) as a function of the bending radius (µm) for both TE and TM modes in H-bend waveguides. The graphs show data at different wavelengths (1.0 µm, 1.5 µm, 2.0 µm, 2.5 µm, and 3.0 µm), highlighting the variations in bending loss for each mode.

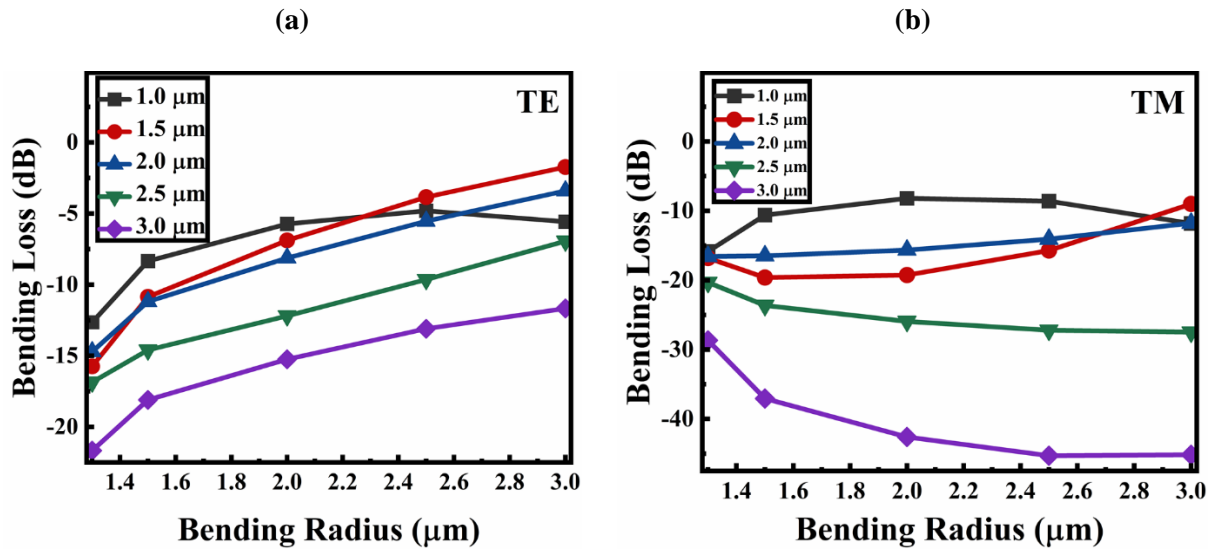

**Figure 8.** Bending Loss analysis at fixed radius 2 um across wavelength (a) TE mode, and (b) TM mode.

Figure 8(a) shows the bending loss as a function of the bending radius for the TE mode at different wavelengths (1.0 µm, 1.5 µm, 2.0 µm, 2.5 µm, and 3.0 µm) in an H-bend waveguide. As the bending radius increases from 1.4 µm to 3.0 µm, a decreasing trend in bending loss is observed for longer wavelengths. At shorter wavelengths such as 1.0 µm and 1.5 µm, the bending loss is initially around -15 dB but decreases significantly as the bending radius increases, resulting in lower loss at larger radii. In contrast, longer wavelengths, such as 2.5 µm and 3.0 µm, experience an even higher initial bending loss that decreases more gradually with an increase in the bending radius. For example, at 1.2 µm bending radius, the loss for the 1.0 µm wavelength is approximately -12 dB, indicating lower attenuation compared to approximately -20 dB loss for the 3.0 µm wavelength. Remarkably, an increase in bending radius results in a reduction of bending loss for all wavelengths, indicating that larger radii improve waveguiding efficiency by minimizing leakage. This behavior highlights the need to optimize the bending radius in H-bend photonic waveguides for efficient light propagation and minimal signal loss. The prime cause of bending losses in waveguides, particularly at smaller bend radii, stems from the misalignment of TE and TM field distributions between the straight and bending sections, leading to inefficient mode coupling and significant energy dissipation [57]. The primary contributor to loss is the mode mismatch at the junction between the straight and bending waveguides, where imperfect alignment of optical modes results in substantial energy loss and diminished transmission efficiency [55]. That is why the bending radius is optimized at 2 μm for further study.

Figure 8(b) displays the bending loss as a factor of the radius of the bending portion for TM mode at different wavelengths (1.0 μm, 1.5 μm, 2.0 μm, 2.5 μm, and 3.0 μm) in an H-bend waveguide. As the bending radius increases from 1.4 μm to 3.0 μm, different wavelengths display varying trends in bending loss. Some wavelengths experience a decrease in bending loss, while others exhibit an increase. At each wavelength (1.0 μm, 1.5 μm, 2.0 μm, 2.5 μm, and 3.0 μm) the bending loss in the H-bend waveguide exhibits distinct behaviors as the bending radius increases from 1.4 μm to 3.0 μm. At shorter wavelengths such as 1.0 μm, 1.5 μm, and 2 μm the bending loss is initially around -15 dB, loss decreases slightly as the bending radius increases, resulting in comparatively lower loss at larger radii. In contrast, longer wavelengths like 2.5 μm and 3.0 μm experience higher initial bending loss than the bending loss at shorter wavelengths, and loss increases gradually with increasing bending radius. For example, at a bend radii of 1.2 μm, the bending loss for the 1.0 μm wavelength is approximately -15 dB, indicating a lower bending loss than approximately -30 dB loss for the 3.0 μm wavelength. At wavelengths of 2.5 μm and 3 μm, the bending loss increases slowly as the bending radius increases, indicating that the loss remains relatively high even with larger bending radii for TE mode. Moreover, the losses for TE (Transverse Electric) polarization are consistently less pronounced than those for TM (Transverse Magnetic) polarization, as demonstrated in Figures 9(a) and (b). This difference arises due to the inherent contrast in refractive indices for in and out-of-plane engagement, which directly impacts the propagation behavior and mode confinement of each polarization, ultimately leading to variations in the magnitude of losses [50]. The principal cause of bending losses in waveguides, particularly at smaller bending radii, stems from the misalignment of dispensation of mode field in the middle of the unbending as well as curve sections, leading to inefficient mode coupling and significant energy dissipation [57]. Similar trends are also seen in some previous reports [58-60]. That is why the bending radius is optimized at 2 µm for further study.

## 4. Conclusion

This study explores the design and simulation of a GaSe-based photonic waveguide on a β-$Ga_2O_3$ platform with an $Al_2O_3$ cladding, demonstrating its potential for advanced photonic integrated circuits. The waveguide exhibits excellent optical confinement, with power confinement factors exceeding 96% for both TE and TM modes at a wavelength of 1.55 µm. The TM mode, in particular, shows lower propagation losses compared to the TE mode, making it more suitable for long-wavelength applications. The analysis based on varying core dimensions and cladding heights reveals that increasing the core height and width enhances optical confinement and reduces propagation losses, with the TM mode consistently outperforming the TE mode in terms of confinement and stability. Additionally, the study highlights the importance of optimizing the bending radius to minimize losses in curved waveguide structures. The GaSe/$Al_2O_3$/β-$Ga_2O_3$ waveguide platform offers a promising alternative to traditional silicon-based photonic circuits, particularly for applications requiring low loss and high efficiency in the near-infrared spectrum. It also paves the way for the integration of GaSe-based waveguides in next-generation photonic devices.

**Corresponding Author:**
[*]E-mail: jak_apee@ru.ac.bd (Jaker Hossain).

**Disclosure:** The authors have no conflicts to disclose.

**Data availability**: Data will be available from the corresponding author upon reasonable request.

**Declaration of generative AI and AI-assisted technologies:** None of the authors use any AI or AI-assisted technologies in writing this manuscript.